\documentclass[twoside,fleqn]{article}
\usepackage{espcrc2}
\usepackage{epsfig}
\newcommand{\plaq}{\mbox{\raisebox{-.75mm}
{\epsfig{file=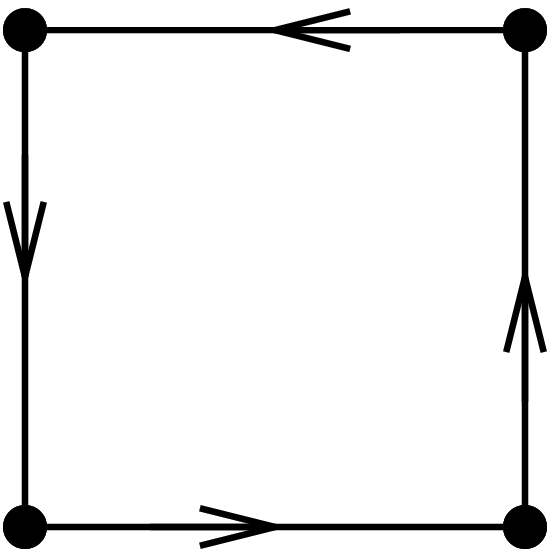,height=3mm
}}~}}
\newcommand{\loopt}{\mbox{\raisebox{-2mm}
{\epsfig{file=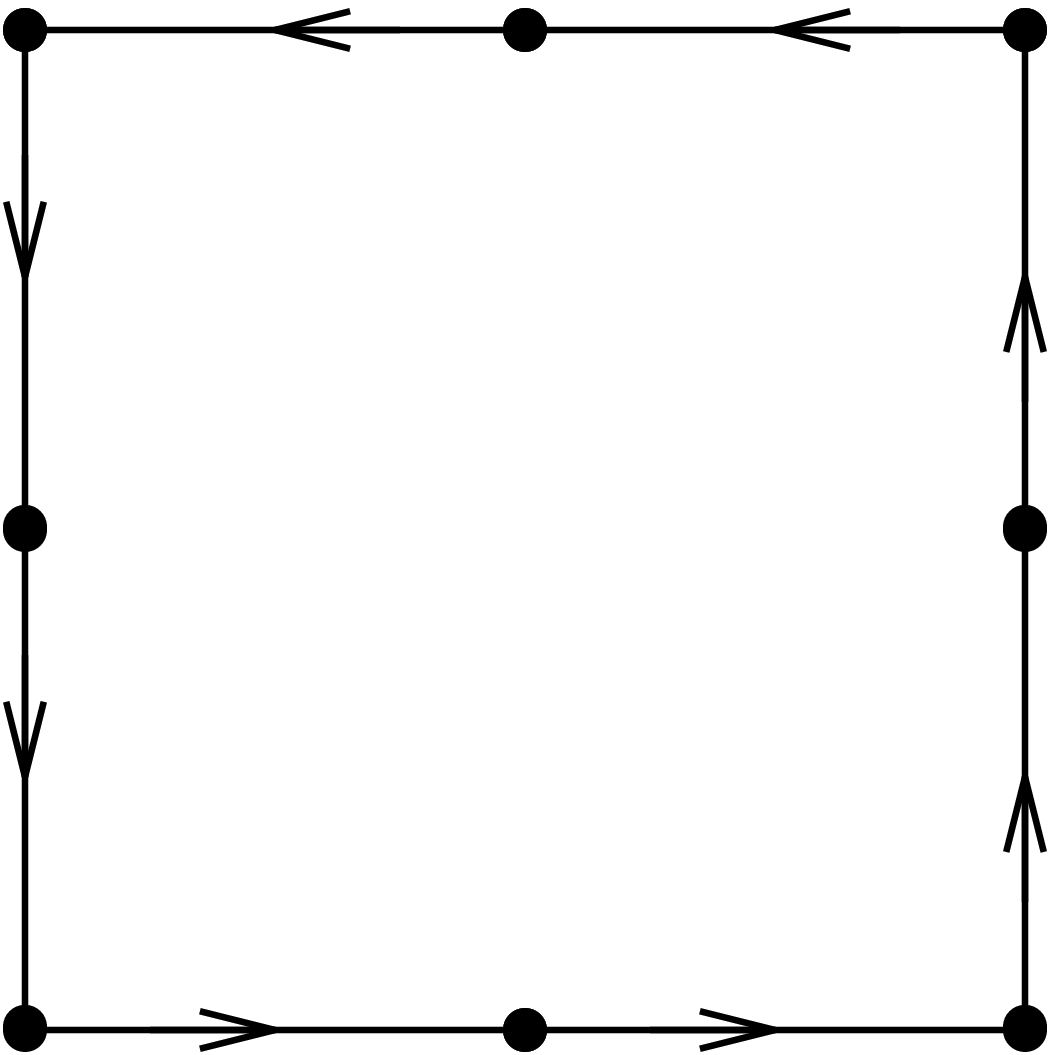,height=6mm
}}~}}
\newcommand{\loOp}{\mbox{\raisebox{-.75mm}
{\epsfig{file=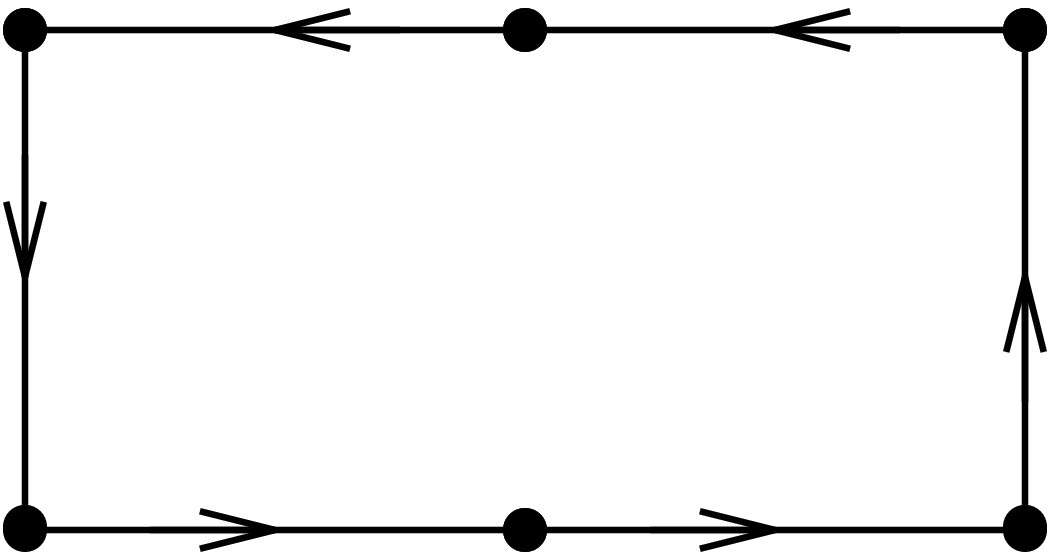,height=3mm
}}~}}
\newcommand{\lOop}{\mbox{\raisebox{-2mm}
{\epsfig{file=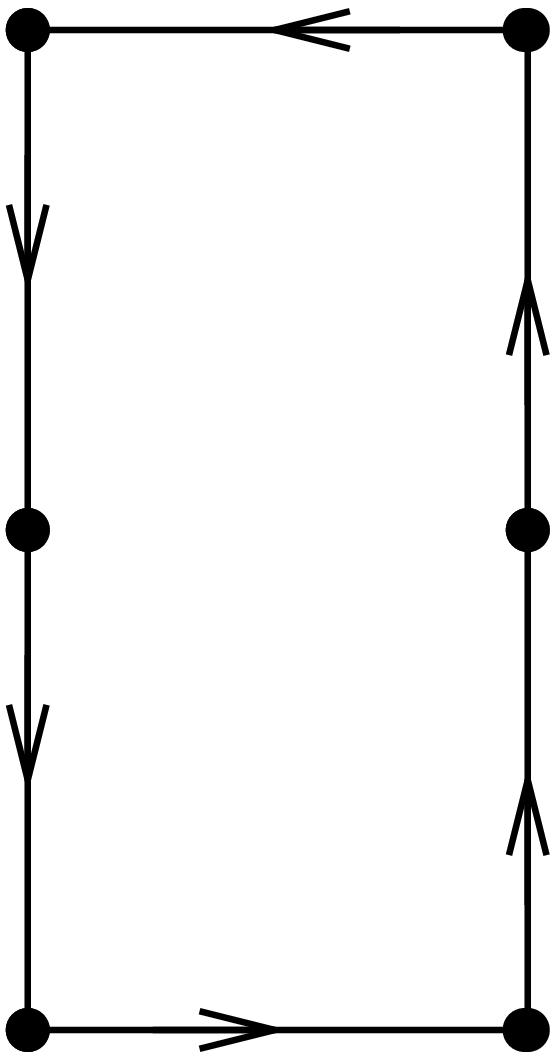,height=6mm
}}~}}
\newcommand{\nn}{\nonumber}
\newcommand{\tr}{\mbox{Tr~}}
\newcommand{\re}{\mbox{Re~}}
\newcommand{\kronecker}[2]{  \delta_{#1 , #2}  }

\newcommand{\mymatrix}[2]{  
        m \kronecker{#1}{#2} + \eta_{#1}\left( \; 
                    \frac{9}{16} A\left[ U \right]_{#1\; #2} - 
        \frac{1}{48} B\left[ U  \right]_{#1\; #2} \right) } 
\newcommand{\linkdagger}[2]{  U_{ #1 , #2}^\dagger  }
\newcommand{\link}[2]{  U_{ #1 , #2}  }
\newcommand{\apart}[3]{      
\sum_{#3} \left( \link{#1}{#3} \kronecker{#1}{#2-\hat{#3}} 
                -\linkdagger{#1-\hat{#3}}{#3} 
                                        \kronecker{#1}{#2+\hat{#3}} \right) }

\newcommand{\bparta}[3]{
\sum_{#3} ( \link{#1}{#3} \link{#1+\hat{#3}}{#3} \link{#1+2 \hat{#3}}{#3} 
                                             \kronecker{#1}{#2- 3 \hat{#3}}   }
\newcommand{\bpartb}[3]{
                 \linkdagger{#1-\hat{#3}}{#3} \linkdagger{#1-2\hat{#3}}{#3} 
                 \linkdagger{#1-3 \hat{#3}}{#3} \kronecker{#1}{#2+ 3 \hat{#3}}
                  ) }

\newcommand{\AmS}{{\protect\the\textfont2
  A\kern-.1667em\lower.5ex\hbox{M}\kern-.125emS}}

\title{
\vskip -100pt
 \mbox{} \hfill BI-TP 96/34\\
 \mbox{} \hfill August 1996\\
\vskip  70pt
QCD Thermodynamics with Improved Actions}

\author{F. Karsch
\thanks{The work has been supported by the DFG under grants 
Pe 340/3-3 and Pe 340/6-1.} 
with B. Beinlich, J. Engels, R. Joswig, E. Laermann, A. Peikert and B.
Petersson \\
\vskip 6pt
Fakult\"at f\"ur Physik, Universit\"at Bielefeld, D-33615 Bielefeld, Germany
}      
\begin{document}

\begin{abstract}
The thermodynamics of the SU(3) gauge theory has been analyzed with tree
level and tadpole improved Symanzik actions. A comparison with the continuum
extrapolated results for the standard Wilson action shows that improved
actions lead to a drastic reduction of finite cut-off effects already on
lattices with temporal extent $N_\tau=4$. Results for the pressure, the
critical temperature, surface tension and latent heat are presented. First
results for the thermodynamics of four-flavour QCD with an improved
staggered action are also presented. They indicate similarly large
improvement factors for bulk thermodynamics. 
\end{abstract}

\maketitle

\section{Why improved actions are expected to improve the QCD thermodynamics}

Although the plasma phase of QCD does in many respects show strong
non-perturbative properties (screening masses, poor convergence of
perturbation theory,...), bulk thermodynamic observables like energy
density and pressure do approach the non-interacting ideal gas limit at
high temperatures ($T$) where they are expected to receive large contributions
from high momentum modes, $\langle {\rm avg. momentum} \rangle
\sim T$. In lattice calculations $T$ is determined through the 
temporal extent $N_\tau$ of the lattice  and the cut-off $a^{-1}$, {\it
i.e.}
$T\equiv 1/N_\tau a$. In the ideal gas limit the relevant momenta are
therefore of the order of the cut-off, where lattice and continuum dispersion 
relations differ strongly from each other. Indeed this is the origin of 
the well-known discrepancy between the energy density of an ideal gas calculated 
on a finite lattice ($\epsilon (N_\tau)$) and in the continuum ($\epsilon_{\rm
SB}$),

\begin{eqnarray}
{\epsilon (N_\tau) \over \epsilon_{\rm SB} }  
\hskip -0.3cm &=& \hskip -0.3cm \cases{  1+
{30  \over 63} \cdot { \pi^2 \over N_\tau^2} +
{\cal O} \bigl( N_\tau^{-4} \bigr) &\hskip -0.2cm , Wilson \cr 
1+ c_I \cdot { \pi^4 \over N_\tau^4} + 
{\cal O} \bigl( N_\tau^{-6} \bigr) &\hskip -0.2cm , 
Symanzik} \nonumber
%\label{wilsongas}
\end{eqnarray}
The standard Wilson action leads to nearly 50\% corrections on a lattice 
with temporal extent $N_\tau=4$ while
the cut-off dependence is drastically reduced in the case of 
Symanzik improved actions which eliminate the leading ${\cal O}(N_\tau^{-2})
\equiv {\cal O}((aT)^2)$ corrections. In the case of the
(1,2)-Symanzik action this is less than 2\% for $N_\tau=4$, {\it i.e.} $c_I
= 0.044$.
This clearly demonstrates the significance of improved actions in the 
ideal gas limit ($T \rightarrow \infty$). 

In a first exploratory analysis we have shown that the improvement found
analytically in the ideal gas limit also persists at finite temperature and
does seem to be important even close to $T_c$ \cite{Bei96}. 
Here we will present further evidence that tree level and tadpole improved 
actions do lead to a significant reduction of finite cut-off effects even 
at $T_c$ \cite{Bei96a,Bei96b,Eng96}.

\section{SU(3) Thermodynamics}

We have analyzed the thermodynamics of the SU(3) gauge theory using
${\cal O}(a^2)$ improved tree level and tadpole improved actions, which have
been constructed by adding to the Wilson pla\-quette action, $S^{(1,1)}$, an
appropriately weighted contribution from planar (1,2) or (2,2) loops,

\begin{eqnarray}
S^{(1,2)}\hskip -0.3cm &=& \hskip -0.3cm 
\sum_{x, \nu > \mu}~ {5\over 3}~\left(
1-\frac{1}{N}\re\tr\plaq_{\mu\nu}(x)\right) \\
& &\hskip -1.6cm -{1\over 6 u_0^2}\left(1-\frac{1}{2N}\re\tr
\left(\loOp_{\mu\nu}(x)+\lOop_{\mu\nu}(x)\right)\right)\nn\\
S^{(2,2)} \hskip -0.3cm &=& \hskip -0.3cm
\sum_{x, \nu > \mu}~ {4\over 3}~\left(
1-\frac{1}{N}\re\tr\plaq_{\mu\nu}(x)\right) \\
& &-{1\over 48}\left(1-\frac{1}{N}\re\tr
\loopt_{\mu\nu}(x)\right)\nn
\label{actions}
\end{eqnarray}

\begin{table*}[hbt]
% space before first and after last column: 1.5pc
% space between columns: 3.0pc (twice the above)
\setlength{\tabcolsep}{1.5pc}
% -----------------------------------------------------
% adapted from TeX book, p. 241
%\newlength{\digitwidth} \settowidth{\digitwidth}{\rm 0}
\catcode`?=\active \def?{\kern\digitwidth}
% -----------------------------------------------------

\caption{Critical temperature in units of $\sqrt{\sigma}$ 
on lattices with temporal extent $N_\tau=4$. Infinite volume
extrapolations of $\beta_c$ for the Symanzik actions are based on lattices 
with $N_\sigma =16$, 24 and 32.}
\label{tab:tc}
\begin{center}
\begin{tabular}{|l|l|l|l|}
\hline
action &
$N_\sigma$&
$\beta_c$&
$T_c/\sqrt{\sigma} $\\ \hline
standard Wilson &$\infty$ &5.69254~(24)&$ 0.5983~(30)$\\
(2,2) Symanzik (tree level) &24&4.3995~(2)&$ 0.624~(4) $\\
(1,2) Symanzik (tree level) &$\infty$&4.07297~(28)&$ 0.631~(3) $\\
(1,2) Symanzik (tadpole)    &$\infty$&4.35228~(39)&$ 0.635~(3) $\\ \hline
\end{tabular}
\end{center}
\end{table*}

\noindent
where the factor $u_0$ appearing in the definition of the (1,2)-Symanzik action 
denotes the tadpole improvement factor defined in terms of the self-consistently 
determined plaquette expectation value \cite{Lep93}.

\subsection{Bulk thermodynamics}

Bulk thermodynamic quantities like the energy density ($\epsilon$) or
pressure ($p$) calculated with the Wilson action on lattices of size
$N_\sigma^3 \times N_\tau$ with $N_\tau = 4$, 6 and 8 \cite{Boy96} 
show a strong cut-off dependence in the plasma phase. 
A first analysis with the (2,2)-Symanzik action \cite{Bei96} has
shown that this cut-off dependence gets drastically reduced already on a
$N_\tau=4$ lattice. The work presented in Ref.~\cite{Bei96} has been
extended in several ways. On lattices with temporal extent $N_\tau=4$ we 
have analyzed now also the
tree-level improved (1,2)-Symanzik action, which shows even smaller cut-off 
distortions in the high temperature limit than the (2,2) action. In addition
we have studied the influence of tadpole improvement in the case of the 
(1,2)-Symanzik action. In all cases we have determined the temperature scale
non-perturbatively through the calculation of the string tension on
symmetric ($N_\sigma^4$) lattices, $T/T_c \equiv \sqrt{\sigma}a(\beta_c)/
\sqrt{\sigma}a(\beta)$. From this calculation we also obtain $T_c$ 
in units of $\sqrt{\sigma}$ for the different actions. These are
given in Table~\ref{tab:tc} together with results for the
Wilson action. In that case $T_c/\sqrt{\sigma}$ has also
been calculated on lattices with $N_\tau$ varying between 4 and 12 and 
extrapolated to the continuum limit. This gave 
$T_c/\sqrt{\sigma} = 0.629\pm 0.003$ \cite {Boy96}. On $N_\tau=4$
lattices the results obtained with improved actions are thus much closer 
to this extrapolated value than those obtained with the Wilson action. 

\begin{figure}[htb]
\vskip -0.8truecm
\vspace{9pt}
   \epsfig{
       file=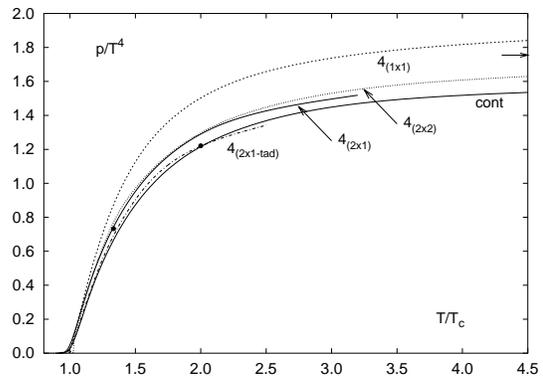,width=75mm}
\vskip -0.7truecm
\caption{Pressure of the SU(3) gauge theory on lattices with temporal extent
$N_\tau = 4$. The solid line shows the continuum extrapolation obtained from
the standard Wilson action. The dots give results from a calculation with a
perfect action on a $12^3\times 3$ lattice [6]. The arrow indicates the
continuum ideal gas value.}
\vskip -0.7truecm
\label{fig:pressure}
\end{figure}

The pressure has been calculated by integrating
differences of the action densities obtained on $16^4$ and $16^3\times 4$
lattices (in the case of the (2,2) action we used $24^4$ and $24^3\times 4$
lattices) following the standard procedures \cite{Bei96,Boy96}. Results are
shown in Figure~\ref{fig:pressure}.

\subsection{Surface tension and latent heat}

The success of improved actions for the calculation of bulk
thermodynamics even at temperatures close to $T_c$ naturally leads to
the question whether these actions also do lead to an improvement at
$T_c$. This is, of course, a highly non-perturbative regime. However,
observables like the latent heat ($\Delta \epsilon$) and the surface 
tension ($\sigma_I$), which characterize the discontinuities at the first 
order deconfinement phase transition in a
$SU(3)$ gauge theory, do depend on properties of the low as well as the
high temperature phase. As the latter is largely controlled by high
momentum modes it may be expected that some improvement does result
even from tree level improved actions.

At the SU(3) deconfinement transition $\Delta \epsilon$ and $\sigma_I$
have been studied on lattices up to temporal extent $N_\tau=6$ 
\cite{Iwa92,Iwa94}. A strong cut-off
dependence has been found when comparing calculations for $N_\tau =4$
and 6. For this reason an extrapolation to the continuum limit 
has so far not been possible for these observables.

\begin{figure}[htb]
\vspace{9pt}
   \epsfig{
       file=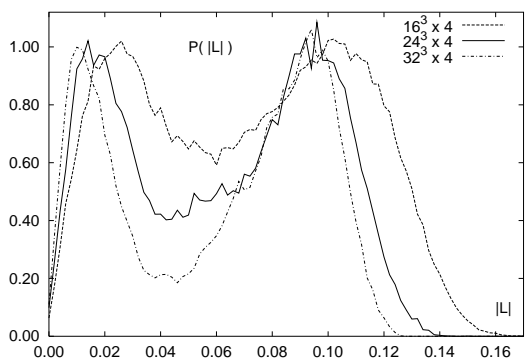,width=75mm}
\vskip -0.7truecm
\caption{Polyakov loop distributions for the tadpole improved (1,2)
Symanzik action. }
\vskip -0.7truecm
\label{fig:ltadpole}
\end{figure}

We have extracted $\sigma_I$ \cite{Bei96b} from the probability 
distribution of the absolute value of the Polyakov loop following the 
analysis presented in Ref.~\cite{Iwa94}. The probability distribution at the
minimum is proportional to
\begin{equation}
P(|L|) \sim \exp\bigl(- \bigl[f_1 V_1 +f_2 V_2 + 2 \sigma_I A\bigr]/T \bigr)
\label{prob}
\end{equation}
where $f_i$ denotes the free energy in the phase $i$, and $V_i$ is the
volume occupied by that phase and $A$ denotes the interface area of the
finite system. From the depth of the minimum one thus can
extract the surface tension. The distribution functions for the tadpole
improved actions for three different lattice sizes are shown in
Figure~\ref{fig:ltadpole}. 

In Table~\ref{tab:surface} we give results for $\sigma_I$ on the largest
lattices considered. Clearly the surface tension extracted
from simulations with improved actions on lattices with temporal extent
$N_\tau=4$ are substantially smaller than corresponding results for the
Wilson action. In fact, they are compatible with the $N_\tau=6$ results for
the Wilson action.

The latent heat is calculated from the discontinuity in 
($\epsilon -3P)$. This in turn is obtained from the
discontinuity in the various Wilson loops entering the definition of
the improved actions,
\begin{eqnarray}
{\Delta\epsilon \over T_c^4} &=&  {1\over 6}
\biggl({N_\tau \over N_\sigma} \biggr)^3 \biggl( a{{\rm d}\beta \over {\rm
d} a}
\biggr) \biggl(\langle \tilde{S} \rangle_+  - \langle \tilde{S} \rangle_-
\biggr) ~,
\label{latent}
\end{eqnarray}
with $\tilde{S} \equiv  S - {\rm d} S / {\rm d} \beta$.

The difference of action expectation values at $\beta_c$ is obtained by
calculating these in the two coexisting phases at 
$\beta_c$. In the time histories of the 
Polyakov loop values we have cuted out the transition regions in order
to classify configurations belonging to either of the two phases
\cite{Iwa92}.  

In order to extract the latent heat one does
still need the $\beta$-function entering the definition of $\Delta
\epsilon /T_c^4$ in Eq.~\ref{latent}. The necessary relation $a(\beta)$ has
been obtained from a calculation of $\sqrt{\sigma}a$ (improved actions) or
a determination of $T_ca$ (Wilson action). 
Results for $\Delta\epsilon/T_c^4$ are summarized in Table~\ref{tab:surface}.
Here we also give the result obtained with the 1-loop $\beta$-function.

\begin{table*}[hbt]
% space before first and after last column: 1.5pc
% space between columns: 3.0pc (twice the above)
\setlength{\tabcolsep}{0.9pc}
% -----------------------------------------------------
% adapted from TeX book, p. 241
%\newlength{\digitwidth} \settowidth{\digitwidth}{\rm 0}
\catcode`?=\active \def?{\kern\digitwidth}
% -----------------------------------------------------
\caption{Surface tension and latent heat for the three improved actions and 
the Wilson action. Results for the Wilson action are based on data from [9] 
using the non-perturbative $\beta$-function calculated in [5].
}
\label{tab:surface}
\begin{center}
\begin{tabular}{|l|c|c|c|c|c|}\hline
action&$V_\sigma$&$N_\tau$&$\sigma_I/T_c^3$&$\Delta\epsilon/T_c^4$ &
$\biggl(\Delta\epsilon/T_c^4\biggr)_{\rm pert}$\\ \hline
standard Wilson &$24^2\times 36$&4&0.0300~(16)&2.27~(5) &4.06~(8)~\\
&$36^2\times 48$&6&0.0164~(26)&1.53~(4) &2.39~(6)~\\ \hline
(1,2) Symanzik (tree level) &$32^3$&4&0.0116~(23)&~1.57~(12) &~2.28~(8)~\\
(1,2) Symanzik (tadpole) &$32^3$&4&0.0125~(17)&~1.40~(9) &~1.94~(8)~\\ \hline
\end{tabular}
\end{center}
\end{table*}

\section{Four-flavour QCD with an improved staggered action}

In the fermionic sector of QCD the influence of a finite cut-off on bulk
thermodynamic observables is known to be even larger than in the pure gauge
sector. For instance, in the staggered formulation the energy density of an 
ideal fermi gas differs by more than 70\% from the continuum value on a
lattice with temporal extent $N_\tau=4$ and approaches the continuum value
only very slowly with increasing $N_\tau$. This cut-off dependence can
drastically be reduced with an $O(a^2)$ improved staggered action. We use
an improved action, $S^I[U] = S^{(1,2)} + \bar{\psi} M \psi$,
%\begin{displaymath}
%\myaction 
%\end{displaymath}
%where the gluonic part is given by the (1,2) Symanzik action and 
where a higher order
difference scheme (one-link and three-link terms), is used to improve the
fermionic part \cite{Nai89}. The improved fermion matrix reads 
\begin{eqnarray*}
M[U]_{ij}\hskip -0.2cm &=&\hskip -0.2cm \mymatrix{i}{j}  \\
{\rm with} & & \\
A[U]_{ij}\hskip -0.2cm 
&=&\hskip -0.1cm \apart{i}{j}{\mu}     \\
%&&\\ 
B[U]_{ij}\hskip -0.2cm &=&\hskip -0.2cm 
\bparta{i}{j}{\mu} - \\
&&\hskip -0.1cm - \bpartb{i}{j}{\mu}
\end{eqnarray*}
With this action the overall cut-off distortion of the ideal gas limit on a
$16^3\times 4$ lattices reduces to about 20\%. We have performed simulations
for two quark masses, $ma =0.05$ and 0.1 \cite{Eng96}. Like in 
the pure gauge sector the improvement is visible already close to $T_c$. For
instance, the energy density stays close to the ideal gas limit. Although we
observe an overshooting of the ideal gas limit close to $T_c$ for the
non-zero quark masses con\-sidered by us this feature does seem to disappear
in the "chiral limit" \cite{Eng96}.
%, which we constructed by eliminating the term
%proportional to $\langle \bar{\psi} \psi \rangle$ in the definition of the
%energy density \cite{Eng96}.

\begin{figure}[htb]
\vskip -0.7truecm
\vspace{9pt}
   \epsfig{bbllx=40,bblly=239,bburx=523,bbury=578,
       file=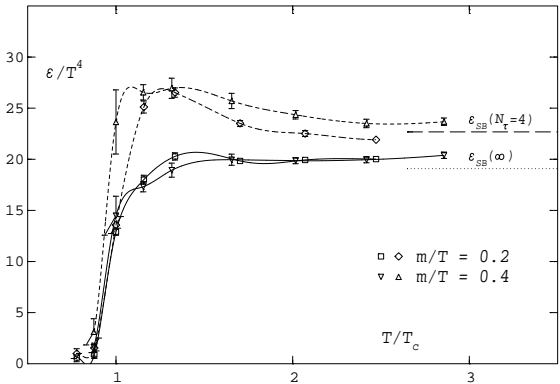,width=75mm}
\vskip -0.7truecm
\caption{Energy density of four-flavour QCD on a $16^3\times 4$ lattice. The
lower set of curves shows an extrapolation to the "chiral limit".}
\label{fig:energy}
\vskip -0.5truecm
\end{figure}

\section{Conclusions}

Thermodynamic observables of SU(3) gauge theory and QCD studied with
improved gauge and fermion actions show a drastic
reduction of the cut-off dependence in the high temperature limit as well
as at $T_c$. The major improvement effect is already obtained with tree
level improved actions. Tadpole improvement plays a minor role even 
close to $T_c$.

\end{document}